\documentstyle[11pt]{article}
\input amssym.def
\input amssym
\input epsf.sty

\newtheorem{theorem}{Theorem}[section]
\newtheorem{lemma}[theorem]{Lemma}
\newtheorem{fact}[theorem]{Fact}

\newtheorem{claim}[theorem]{Claim}
\newtheorem{corollary}[theorem]{Corollary}

\newenvironment{proof}{{\bf Proof:  }}{\qquad\rule{2mm}{2mm}}
\newenvironment{proofof}[1]{{\bf Proof of #1:  }}{\qquad\rule{2mm}{2mm}}

\newcommand{\union}{\;\cup\;}
\newcommand{\intersect}{\;\cap\;}
\newcommand\meet\wedge
\newcommand\implies\Rightarrow

\newcommand{\prob}[1]{\mbox{{\bf Pr}$\left[#1\right]$}}
\newcommand{\size}[1]{\left|#1\right|}
\newcommand{\ceil}[1]{\left\lceil#1\right\rceil}
\newcommand{\floor}[1]{\left\lfloor#1\right\rfloor}

\newcommand{\norm}[1]{\left\|\,#1\,\right\|}
\newcommand{\set}[1]{{\left\{#1\right\}}}
\newcommand{\st}{{\; : \;}}

\newcommand{\ignore}[1]{}
\newcommand{\eq}{\;=\;}

\newcommand{\qq}{\hat{q}}
\newcommand{\aaa}{{\cal A}}
\newcommand{\rank}{{\rm rank}}

\begin{document}

\title{\bf The quantum query complexity of approximating the
           median and related statistics \thanks{Part of this work was
	   done when the first author was at the 1998 Elsag-Bailey --
	   I.S.I. Foundation research meeting on quantum computation.} }
\author{
{\em Ashwin Nayak} \thanks{Computer Science Division, UC~Berkeley.
Email: {\tt ashwin@cs.berkeley.edu}.  Supported by JSEP grant FDF
49620-97-1-0220-03-98.}
   \and
{\em Felix Wu} \thanks{Computer Science Division, UC~Berkeley.
Email: {\tt felix@cs.berkeley.edu}. Supported by an NDSEG Fellowship.}
}
\date{}

\maketitle

\begin{abstract}\noindent\normalsize%
Let~$X = (x_0,\ldots,x_{n-1})$ be a sequence of~$n$ numbers.
For~$\epsilon > 0$, we say that~$x_i$ is an~$\epsilon$-approximate
median if the number of elements strictly less than~$x_i$, and
the number of elements strictly greater than~$x_i$ are each 
less than~$(1+\epsilon)n/2$. We consider the quantum query complexity of
computing an~$\epsilon$-approximate median, given the sequence~$X$ as
an oracle. We prove a lower bound of~$\Omega(\min\{{1\over\epsilon},n\})$
queries for any quantum algorithm that computes an~$\epsilon$-approximate
median with any constant probability greater than~$1/2$.
We also show how an~$\epsilon$-approximate median may be computed
with~$O({1\over\epsilon}\log({1\over\epsilon})
\log\log({1\over\epsilon}))$ oracle queries, which represents an 
improvement over an earlier
algorithm due to Grover~\cite{gr-median,gr-framework}.
Thus, the lower bound we obtain is essentially optimal. The upper
and the lower bound both hold in the comparison tree model as well.

Our lower bound result is an application of the polynomial paradigm recently
introduced to quantum complexity theory by Beals {\em et
al.}~\cite{bbcmw}. The main ingredient in the proof is a
polynomial degree lower bound for real multilinear polynomials that
``approximate'' symmetric {\em partial\/} boolean functions. The degree
bound extends a result of Paturi~\cite{pa} and also
immediately yields lower bounds for the problems of
approximating the~$k$th-smallest element, approximating the mean of a
sequence of numbers, and that of approximately 
counting the number of ones of a
boolean function.  All bounds obtained come within polylogarithmic
factors of the optimal (as we show by presenting algorithms where no
such optimal or near optimal algorithms were known),
thus demonstrating the power of the polynomial method.
\end{abstract}

\section{Introduction}

\subsection{Synopsis}

Proving non-trivial lower bounds for any universal model of computation
is a formidable task, and quantum computers are no exception to this. It is
thus natural to seek bounds in restricted settings. The first such step
in the field of quantum computation was taken by Bennett {\em et
al.}~\cite{bbbv}. They prove that we cannot solve NP-complete problems
in sub-exponential time on a quantum computer merely by adopting the
brute-force strategy of ``guessing'' solutions and checking them for
correctness. Nonetheless, Grover's search algorithm~\cite{gr-search}
shows that
a {\em quadratic\/} speed-up over classical algorithms is possible in
this case. Thus, while the parallelism and the potential for interference
inherent in quantum computation are not sufficient to 
significantly speed up certain strategies for solving problems, they {\em
do\/} give some advantage over probabilistic computation. These results
motivate the question as to whether similar speed up is possible in
other scenarios as well.

Strategies such as `brute-force search' may formally
be modelled via ``black-box'' computations, in which information about the
input is supplied to the algorithm by an {\em oracle}. For example,
the black-box search problem may be defined as follows: given oracle access
to~$n$ bits~$X = (x_0, \ldots, x_{n-1})$, compute
an index~$i$ such that~$x_i = 1$, if such an index exists. A simpler
formulation would require a yes/no answer according to whether such
an index exists or not. This amounts to computing the logical OR of the
input bits. In the black-box setting, strategies  are evaluated by
studying the {\em query complexity\/} of the problem, i.e., the 
minimum, over all algorithms,
of number of times the oracle is accessed (in the worst case) to solve
the problem.
In the case of the abstract search problem,
the query complexity is the number of bits that need to be examined
(in the worst case) in order to compute the logical OR of the~$n$ bits.

Considerable success has been achieved in the study of the query
complexity of computing {\em boolean functions\/} in the quantum
black box model, both in terms of optimal lower bounds for specific
functions~\cite{bbbv,bbht,fggs,bbcmw}, and in terms of general
techniques for proving such lower bounds~\cite{bbbv,bcw,bbcmw}.
However, few approaches were known for the
study of more general functions. Consider, for example, the problem of
approximating the median of~$n$ numbers. An~{\it
$\epsilon$-approximate median\/} of a sequence~$X = (x_0, \ldots,
x_{n-1})$ of~$n$ numbers is a number~$x_i$ such that the number of~$x_j$
less than it, and the number of~$x_j$ more than it are both
less than~$(1+\epsilon){n\over 2}$.
The problem then is to compute such an~$x_i$, given, as an oracle, the
sequence~$X$ of input values, and an explicitly specified
parameter~$\epsilon > 0$ (which may be assumed to be at least~$1\over {2n}$).
Grover gave an algorithm for finding
an~$\epsilon$-approximate median that
makes~$\tilde{O}({1 \over \epsilon})$ queries to the input
oracle~\cite{gr-median,gr-framework}. (Here, the~$\tilde{O}$ notation
suppresses factors involving~$\log({1 \over \epsilon})$ and~$M$,
where~$M$ is the size of the domain the numbers are picked from.) Thus,
an almost quadratic speed up over the best classical algorithm was achieved
(assuming~$M$ to be constant). However, it was still open whether this
algorithm could be improved upon. In particular, known 
techniques such as the ``hybrid argument'' yielded a lower bound
of~$\Omega({1 \over \sqrt{\epsilon}})$ for the number of
queries~\cite{va},
whereas~$O({1 \over \epsilon})$ was suspected to be optimal. In this
paper, we prove a lower bound  of~$\Omega({1 \over \epsilon})$ for the
query complexity of the approximate median problem, thus showing that
Grover's algorithm is almost optimal.
We also present a new~$O({1\over\epsilon} \log({1\over\epsilon})
\log\log({1\over\epsilon}))$ query algorithm for the problem,
thereby eliminating the dependence of the upper bound on~$M$.
The upper and the lower
bound both also hold in the {\em comparison tree model}, in which one is
interested in the number of {\em comparisons between the input
elements\/} required to compute an~$\epsilon$-approximate median.

Our lower bound is derived via the {\em polynomial method\/} recently
introduced to the area of quantum computing by Beals {\em et
al.}~\cite{bbcmw}. They
show that the acceptance probability of
a quantum algorithm making~$T$ queries to a boolean oracle can be expressed as
a real multilinear polynomial of degree at most~$2T$ in the oracle
input. Thus, if the algorithm computes a boolean function of the oracle
input with probability at least~$2/3$, the polynomial {\em approximates\/}
the function to within~$1/3$ at all points in the boolean hypercube. So, by
proving a lower bound on the degree of polynomials approximating the boolean
function, we can derive a lower bound on the number of queries~$T$
the quantum algorithm makes. We cannot, however, follow this particular
route for the problem of approximating the median, since the
restriction of the problem to boolean inputs does not yield a
well-defined function. Nonetheless, the restriction {\em does\/} yield a
{\em partial\/} boolean function, i.e., a function that is not defined at all
points of the domain. Our result is thus based on a degree lower bound
for polynomials that ``approximate'' partial boolean functions. This
degree lower bound generalizes a bound due to Paturi~\cite{pa},
and also gives lower bounds for the problems of approximating the~$k$th
smallest element, approximating the mean of a sequence of numbers, and
that of approximately counting the number of ones
of a boolean function. All bounds
obtained are almost tight (as we show by presenting algorithms where no
such optimal or near optimal algorithms were know),
demonstrating the power of the polynomial method.

\subsection{Summary of results}
\label{sec-summ}

Consider a partial boolean function~$f : \set{0,1}^n \rightarrow
\set{0,1}$. We say a real~$n$-variate polynomial~$p$ {\em
approximates\/} the partial function~$f$ to within~$c$, for a
constant~$0 \le c < 1/2$, if
\begin{enumerate}
\item for all~$X \in \set{0,1}^n$,~$p(X) \in [-c, 1+c]$, and
\item for all points~$X$ at which~$f$ is defined,~$\size{p(X) - f(X)}
      \le c$.
\end{enumerate}
Our main theorem gives a degree lower bound for polynomials
approximating partial boolean functions of the following type. For~$X
= (x_0,\ldots,x_{n-1}) \in \set{0,1}^n$, let~$\size{X} =
\sum_{i = 0}^{n-1} x_i$ be the number of ones in~$X$.
Further, let~$\ell,\ell'$ be integers such that~$0 \le
\ell\not=\ell' \le n$. Define the partial boolean
function~$f_{\ell,\ell'}$ on~$\set{0,1}^n$ as
\begin{eqnarray*}
f_{\ell,\ell'}(X) & = & \left\{ \begin{array}{ll}
                             1 & \mbox{if }\size{X} = \ell \\
                             0 & \mbox{if }\size{X} = \ell' \\
                         \end{array}
                 \right.
\end{eqnarray*}
Let~$m \in \set{\ell,\ell'}$ be such that~$\size{{n\over 2}-m}$ is
maximized, and let~$\Delta_\ell = \size{\ell-\ell'}$.
\begin{theorem}
\label{thm-main}
Let~$p$ be any real~$n$-variate polynomial which approximates the
partial boolean function~$f_{\ell,\ell'}$ to within~$c$,
for some constant~$c < 1/2$.
Then, the degree of~$p$ is~$\Omega(\sqrt{n/\Delta_\ell}
+ \sqrt{m(n-m)}/\Delta_\ell)$.
\end{theorem}
This theorem subsumes a degree lower bound given by Paturi~\cite{pa} for
polynomials approximating (total) symmetric boolean functions.

We say that an algorithm~$\cal A$, possibly with access to an oracle, 
{\em computes\/} a {\em partial\/} function~$f$ on~$\set{0,1}^n$,
if~$\prob{{\cal A}(X) \not= f(X)} \le \delta$ for all inputs~$X$ for
which~$f$ is defined, where~$\delta$ is some constant less than~$1/2$.
For boolean~$f$, we say that the algorithm {\em accepts\/} an input~$X$
if~${\cal A}(X) = 1$. Theorem~\ref{thm-main}, when combined with a
characterization due to Beals {\it et al.\/} (Lemma~4.2 of~\cite{bbcmw}) 
of the probability of acceptance of a quantum algorithm on a
boolean input oracle, in terms of
polynomials, gives us the following result.
\begin{corollary}
\label{thm-q-main}
Any quantum black-box algorithm that computes the partial boolean
function~$f_{\ell,\ell'}$, given the input as an oracle, 
makes~$\Omega(\sqrt{n/\Delta_\ell} + \sqrt{m(n-m)}/\Delta_\ell)$ queries.
\end{corollary}
This lower bound also holds for the {\em expected\/} query complexity of
computing the partial function~$f_{\ell,\ell'}$.
Using an approximate counting algorithm of Brassard {\em et
al.}~\cite{bht,mo,bhmt}, we show that our query
lower bound is optimal to within a constant factor.
\begin{theorem}
\label{thm-q-ub}
The quantum query complexity of computing the partial
function~$f_{\ell,\ell'}$, given the input as an oracle,
is~$O(\sqrt{n/\Delta_\ell} + \sqrt{m(n-m)}/\Delta_\ell)$.
\end{theorem}
The result of Beals {\em et al.\/} mentioned above
then immediately implies that the degree lower bound of
Theorem~\ref{thm-main} is also optimal to within a constant factor.
\begin{corollary}
\label{thm-main-ub}
For any constant~$0 < c < 1/2$, there is a real,~$n$-variate polynomial~$p$ of
degree~$O(\sqrt{n/\Delta_\ell} + \sqrt{m(n-m)}/\Delta_\ell)$ that
approximates the function~$f_{\ell,\ell'}$ to within~$c$.
\end{corollary}

Corollary~\ref{thm-q-main} enables us to prove lower bounds
for the query complexity of computing the statistics listed below,
given, as an oracle, a list~$X = (x_0, \ldots, x_{n-1})$ of (rational)
numbers in the range~$[0,1]$ and an explicitly
specified real parameter~$\epsilon > 0$ or~$\Delta > 0$.
We may assume~$\epsilon$ to be in the range~$[1/(2n),1)$, and~$\Delta$ to
be in~$[1/2, n)$.
\begin{enumerate}

\item {\bf $\epsilon$-approximate median}. A number~$x_i$
such that~$\size{\set{j \st x_j < x_i}} < (1+\epsilon)n/2$ and~$
\size{\set{j \st x_j > x_i}} < (1+\epsilon)n/2$.

\item {\bf $\Delta$-approximate $k$th-smallest element}. (Defined
for~$1 \le k \le n$.)
A number~$x_i$ that is a~$j$th-smallest element of~$X$ for some~$j$
in the range~$(k-\Delta,k+\Delta)$.

\item {\bf $\epsilon$-approximate mean}. A number~$\mu$ such
that~$\size{\mu - \mu_X} < \epsilon$,
where~$\mu_X = {1 \over n} \sum_{i = 0}^n x_i$ is the
mean of the~$n$ input numbers.

\item {\bf $\Delta$-approximate count}. (Defined when~$x_i \in \set{0,1}$ for
all~$i$.) A number~$t$ such that~$\size{t - t_X} < \Delta$, where~$t_X
= \size{X} = \sum_{i = 0}^n x_i$ is the number of ones in~$X$.

\item {\bf $\epsilon$-approximate relative count}. (Defined when~$x_i
\in \set{0,1}$ for all~$i$.) A number~$t$ such that~$\size{t - t_X} <
\epsilon t_X$, where~$t_X$ is defined as above.

\end{enumerate}
Note that some of the problems defined above are very
closely related to each other.
Problem~2 is a natural generalization of problem~1;
problem~4 is, of course, the restriction of problem~3 to boolean inputs
(with~$\Delta$ defined appropriately), and problem~5 is a version of
problem~4 where we are interested in bounding {\em relative\/}
error rather than {\em additive\/} error. In the case of problems~1
and~2, we may relax the condition that the approximate statistic be a
number from the input list (with a suitable modification to definition~2
above); our results continue to hold with the relaxed definitions.
(Problem~1 was first studied by Grover~\cite{gr-median,gr-framework}
with this relaxed definition.)

We first prove a lower bound for approximating the~$k$th-smallest
element by showing reductions from partial functions of the sort
described above.
\begin{theorem}
\label{thm-kth-lb}
At least~$\Omega(\sqrt{n/\Delta}+ \sqrt{k(n-k)}/\Delta)$ oracle 
queries are made by any quantum black-box algorithm for computing
a~$\Delta$-approximate~$k$th-smallest element.
\end{theorem}
We thus get a lower bound for the approximate median problem as well.
\begin{corollary}
\label{thm-median-lb}
The quantum query complexity of computing an~$\epsilon$-approximate
median is~$\Omega(1/\epsilon)$.
\end{corollary}
We also propose an algorithm for approximating the~$k$th-smallest
element that comes within a polylogarithmic factor of the optimum.
\begin{theorem}
\label{thm-kth-ub}
Let~$N = \sqrt{n/\Delta} + \sqrt{k(n-k)}/\Delta$. There is a quantum
black-box algorithm that computes a~$\Delta$-approximate $k$th-smallest
element of~$n$ numbers given via an oracle, with~$O(N\log(N)
\log\log(N))$ queries.
\end{theorem}
This gives us a new algorithm for estimating the median.
Our algorithm represents an improvement over the algorithm of
Grover~\cite{gr-median,gr-framework} when the input numbers are allowed
to be drawn from an arbitrarily large domain.
\begin{corollary}
\label{thm-median-ub}
$O({1\over\epsilon} \log({1\over\epsilon}) \log\log({1\over\epsilon}))$
queries are sufficient for computing an~$\epsilon$-approximate median in
the black-box model.
\end{corollary}
This gives us an almost quadratic speed up over classical algorithms in
the worst case.

A very natural measure of complexity of computing functions such as
the~$k$th-smallest element of a given list of numbers is the number of
{\em comparisons\/} between the input elements required for the
computation. To study this aspect of such problems, one considers
algorithms in the {\em comparison tree model}. 
In this model, the algorithm is
provided with an oracle that returns the result of the comparison~$x_i <
x_j$ when given a pair of indices~$(i,j)$, rather than an oracle that
returns the number~$x_i$ on a query~$i$, where the~$x_i$'s are
understood to be the input numbers. The query complexity of a
problem such as computing the minimum or the median 
then exactly corresponds to the number of {\em comparisons\/}
required to solve the problem. The lower and the upper bounds given above
for estimating the~$k$th-smallest element and the median
continue to hold in the comparison tree model. In particular,
if we set~$\Delta = 1$, we get an almost
optimal~$\tilde{O}(\sqrt{k(n-k+1)}\,)$
comparison algorithm for computing the~$k$th-smallest element (c.f.
Theorems~\ref{thm-kth-lb} and~\ref{thm-kth-ub}).
(An optimal~$O(\sqrt{n})$ comparison algorithm was already known for
computing the minimum of~$n$ numbers~\cite{dh}.)
This should be contrasted with the bound of~$\Theta(n)$ in the classical
case~\cite{bfprt}.
\begin{corollary} Let~$N = \sqrt{k(n-k+1)}$.
Any comparison tree quantum algorithm that computes
the~$k$th-smallest element of a list of~$n$ numbers
makes~$\Omega(N)$ comparisons. Moreover, there is a
quantum algorithm that solves this problem with~$O(N\log(N)
\log\log(N))$ comparisons.
\end{corollary}

Another application of Corollary~\ref{thm-q-main} is to the problem of
approximating the mean. Grover~\cite{gr-framework} recently gave
an~$O({1\over\epsilon} \log\log{1\over\epsilon})$ query algorithm for
this problem, which is again an almost quadratic improvement over classical
algorithms. When the inputs are restricted to be~0/1, the problem
reduces to the counting
problem. Using the approximate counting algorithm of Brassard {\it et
al.\/} mentioned above, we show that the computation of the mean can
be made sensitive to the number of ones in the input, thus getting
better bounds when~$ \size{t-n/2}$ is large.
\begin{theorem}
\label{thm-count-ub}
There is a quantum black-box algorithm that, given a boolean
oracle input~$X$, and an integer~$\Delta > 0$, computes
a~$\Delta$-approximate count and makes an expected~$O(\sqrt{n/\Delta} +
\sqrt{t(n-t)}/\Delta)$ number of queries on inputs with~$t$ ones.
\end{theorem}
We show that this algorithm is optimal to within a constant factor,
and, in the process, get an almost tight lower bound for the general
mean estimation problem.
\begin{theorem}
\label{thm-count-lb}
Any quantum back-box algorithm that approximates the number of ones of a
boolean oracle to within an additive error of~$\Delta$ 
makes~$\Omega(\sqrt{n/\Delta} + \sqrt{t(n-t)}/\Delta)$ queries on
inputs with~$t$ ones.
\end{theorem}
\begin{corollary}
\label{thm-mean-lb}
The quantum query complexity of the~$\epsilon$-approximate mean problem
is~$\Omega({1\over\epsilon})$.
\end{corollary}
Brassard {\it et al.}~\cite{bht,mo,bhmt} study the version of the
approximate counting problem in which one is interested in
bounding the {\em relative\/} error of the estimate. We show that their
algorithm is optimal to within a constant factor (when~$t \le
(1-\epsilon) n$).
\begin{theorem}
\label{thm-relcount-lb}
Any quantum back-box algorithm that solves the~$\epsilon$-approximate 
relative count problem makes
$$\Omega\left(\sqrt{\frac{n}{\ceil{\epsilon(t+1)}}} + \frac{
\sqrt{t(n-t)}}{ \ceil{\epsilon(t+1)}} \right) $$
queries on inputs with~$t$ ones.
\end{theorem}

Finally, we would like to point out that in view of
Corollary~\ref{thm-main-ub}, the lower bounds stated above cannot be
improved using the method we employ in this paper. In fact, we believe
that the lower bounds are optimal, and that the upper bounds can be
improved to match them (up to constant factors).

\section{The lower bound theorem and its applications}

This section is devoted to deriving a polynomial degree lower bound, and
to showing how lower bounds for the query
complexity of the different black-box problems defined in
Section~\ref{sec-summ} follow from it. We first prove the degree lower
bound for polynomials in Section~\ref{sec-main}, and then apply the result
to quantum black-box computation in Section~\ref{sec-apps}.

\subsection{A degree lower bound for polynomials}
\label{sec-main}

We now prove our main result, Theorem~\ref{thm-main}, which  gives
a lower bound for polynomials
approximating symmetric partial functions. The bound is derived using a
technique employed by Paturi~\cite{pa} for polynomials that approximate
non-constant symmetric boolean functions. Our bound generalizes and subsumes
the Paturi bound.

We refer the reader to Appendix~\ref{sec-prop} for the definition of the
concepts involved in the proof. Appendix~\ref{sec-prop} also summarizes
the various facts about polynomials that we use to derive the bound. 

Our proof rests heavily on the inequalities of
Bernstein and Markov (Facts~\ref{thm-bern} and~\ref{thm-bern-mark}).
The essence of these inequalities is that if there is a point in~$[-1,1]$ at
which a polynomial has a ``large'' derivative, and if the point is suitably
close to the middle of the interval, the polynomial has ``high'' degree.

\begin{proofof}{Theorem~\ref{thm-main}}
Recall from Section~\ref{sec-summ} that~$f_{\ell,\ell'}(X)$ is a partial
boolean function on~$\set{0,1}^n$
which is~$1$ when~$\size{X} = \ell$ and~$0$
when~$\size{X} = \ell'$, that~$m$ is one of the integers~$\ell,\ell'$
such that~$\size{n/2-m}$ is maximized, and
that~$\Delta_\ell = \size{\ell-\ell'}$.
We assume that~$p$ is an $n$-variate polynomial of degree~$d$
which approximates the partial function~$f$
to within~$1/3$ in the sense defined in Section~\ref{sec-summ}.
The constant~$1/3$ may be replaced by any constant less than~$1/2$; the proof
continues to hold for that case.
Without loss of generality, we assume that~$\ell > \ell'$ 
(we work with the polynomial~$1-p$, which approximates~$1-f$, 
if~$\ell < \ell'$).

We begin by replacing~$p$ with its {\em symmetrization\/}~$p^{\rm sym}$
and then using Fact~\ref{thm-sym} to transform it to an
equivalent {\em univariate\/} polynomial~$q$. (Since~$x^2 = x$ for~$x
\in \set{0,1}$, we may assume that~$p$ is {\em multilinear}.) We show a degree
lower bound for~$q$, thus giving a degree lower bound for~$p$.

In order to apply the derivative inequalities above, we scale to
transform the
polynomial~$q$ to an equivalent~polynomial $\qq$ over the interval~$[-1,1]$,
where~$\qq(x) = q((1+x)n/2)$. For~$i= 0,1,\ldots,n$,
let~$a_i = 2i/n - 1$. Clearly,~$\qq$ has the following properties:
\begin{enumerate}
\item $\qq$ has degree at most~$d$.
\item $\size{\qq(a_i)} \leq 4/3$ for~$0\le i \le n$.
\item $\qq(a_\ell) \geq 2/3$ and~$\qq(a_{\ell'}) \leq 1/3$. Thus, by the
Mean Value Theorem, there is a point~$a$ in the
interval~$[a_{\ell'},a_\ell]$ such that~$\qq'(a) \geq
(2/3-1/3)/(a_\ell-a_{\ell'}) = n/(6\Delta_\ell)$.
\end{enumerate}

We prove two lower bounds for~$d$, which together imply the theorem.
The first of the lower bounds follows by applying the Markov Inequality
(Fact~\ref{thm-bern-mark}.1) directly to~$\qq$.
\begin{lemma}
\label{thm-first}
$d = \Omega(\sqrt{n/\Delta_\ell}\,)$.
\end{lemma}
\begin{proof}
We consider two cases:

{\bf Case~(a).} $\norm{\qq} < 2$. Combining property~$3$ of~$\qq$ listed
above and Fact~\ref{thm-bern-mark}.1, we get
$$d^2 \;\geq \;\qq'(a)/\norm{\qq} \;\ge\; n/(12\Delta_\ell).$$
So~$d = \Omega(\sqrt{n/\Delta_\ell}\,)$.

{\bf Case~(b).} $\norm{\qq} \ge 2$. From property~$2$ of~$\qq$ listed
above, every point at which~$\qq$
attains its norm is no more than~$2/n$ away from a point~$a_i$ at
which~$\size{\qq(x)} \leq 4/3$. Hence, by the Mean Value Theorem, 
there is a point~$\hat{a} \in
[-1,1]$ such that
$$\size{\qq'(\hat{a})} \;\geq \; (\norm{\qq}-4/3)/(2/n) \;\ge\;
n\norm{\qq}/6.$$
The Markov inequality then implies~$ d = \Omega(\sqrt{n}\,) =
\Omega(\sqrt{n/\Delta_\ell}\,)$.
\end{proof}

The second of the lower bounds now follows from an application of
the Bernstein Inequality for algebraic and {\em trigonometric\/}
polynomials (Facts~\ref{thm-bern-mark}.2 and~\ref{thm-bern},
respectively).
\begin{lemma}
\label{thm-second}
$d = \Omega(\sqrt{m(n-m)}/\Delta_\ell)$.
\end{lemma}
\begin{proof}
Note that if~$\qq$ has norm less than~$2$,
property~3 in conjuntion with Fact~\ref{thm-bern-mark}.2 implies
that
$$ 2d \;\ge\; \norm{\qq}d \;\ge\; \sqrt{1-a^2}\,\qq'(a) \;\ge\; \sqrt{1-a^2} \,
(n/6\Delta_\ell).$$
But since~$a \in [a_{\ell'}, a_\ell]$, we have
$$ 1-a^2 \;\ge\; 1-a_m^2 \eq 1-(2m/n-1)^2 \eq
4m(n-m)/n^2.$$
So,~$d = \Omega(\sqrt{m(n-m)}/\Delta_\ell)$.

Now suppose that~$\norm{\qq} \ge 2$. The proof in this case 
is not as straightforward as
in Case~(b) of the proof of Lemma~\ref{thm-first}, since
Fact~\ref{thm-bern-mark}.2 only gives us a bound which is
sensitive to the point at
which~$\qq$ has high derivative. However, it is possible to ``damp''
the value of the polynomial outside a suitable interval, and thus obtain
the required bound.

Let~$b$ be a point in~$[-1,1]$ at which~$\min_x
\set{ \size{x} \st \size{\qq(x)} \ge 2 }$ is attained, and let~$c$ be one
of the numbers~$b,a_\ell$ such that~$\size{c}$ is minimized.
We assume that~$c \ge 0$, since the proof in the other case is similar.
Let~$C$ be a constant such that~$0 < C < 0.01$.
We distinguish between two cases.

{\bf Case~(a).} $c \le 1-C$. We consider a polynomial~$r$ defined as:
$$ r(x) \; = \; \qq(x+c)(1-x^2)^{d_1} $$
where~$d_1 = \ceil{6/C^2}d$. The degree~$D$ of~$r$ is clearly~$O(d)$, so it
suffices to prove the claimed lower bound for~$D$.

Suppose~$\norm{r} < 2$. Then, the following property of~$r$ gives us
the required bound.
If~$c = a_\ell$, then~$r(0) \ge 2/3$, and we also have~$r(a_{\ell'}-c)
\le 1/3$. If~$c = b$, then~$\size{r(0)} \ge 2$,
and moreover, there is a point~$\hat{c} <
c$ at a distance at most~$2/n$ from~$c$ such that~$\size{r(\hat{c}-c)} 
\le 4/3$. In
either case, there is a point~$\hat{a}\in [a_{\ell'}-a_\ell,0]$ such
that~$\size{r'(\hat{a})} = \Omega(n/\Delta_\ell)$. We may assume,
without loss of generality, that~$\Delta_\ell \le n/4$,
so that~$\hat{a}\in [-1/2,0]$. (Indeed, since~$d \ge 1$, we already
have~$d = \Omega(\sqrt{m(n-m)}/\Delta_\ell)$, if~$\Delta_\ell > n/4$).
We may now invoke the Mean Value Theorem and Fact~\ref{thm-bern-mark}.2
to conclude that~$D =
\Omega(n/\Delta_\ell) = \Omega(\sqrt{m(n-m)}/\Delta_\ell)$.

We now focus on the case when~$\norm{r} \ge 2$.
We show in Claim~\ref{thm-bd1} below that~$\size{r(x)}$ is bounded by~$1$
for~$C \le \size{x} \le 1$. This implies that~$\norm{r}$ (which
is at least~$2$) is
attained within~$[-C,C]$. Note that~$r$ is bounded by~$4/3$
at points~$a_i-c$ separated by at most~$2/n$
in~$[-C,C]$. So there is a point~$\hat{a}
\in [-C,C]$ at which~$\size{r'(\hat{a})} \ge n\norm{r}/6$.
Applying Fact~\ref{thm-bern-mark}.2 to~$r$ at the point~$\hat{a}$,
we get~$D = \Omega(n) = \Omega(\sqrt{m(n-m)}/\Delta_\ell)$.

It only remains to prove the following claim
to complete the analysis of Case~(a).
\begin{claim}
\label{thm-bd1}
For all~$x \in [-1,-C]\union[C,1]$, we have~$\size{r(x)} \le 1$. 
\end{claim}
\begin{proof}
Note that~$\norm{\qq} = \max_{0\le x\le n} \size{q(x)}$. By
Fact~\ref{thm-normbd}, we thus have~$\norm{\qq} \le (4/3)\cdot 2^d$.
In particular,~$\size{\qq(x+c)} \le (4/3)\cdot 2^d \le (4/3)
\cdot {\rm e}^{5d}$ for~$x \in [-1,1-c]$.
We give the same bound for~$\size{\qq(x+c)}$
for~$x \in [1-c,1]$ by using Fact~\ref{thm-chebbd}:
$$ \size{\qq(x+c)} \;\le\; \norm{\qq} \cdot T_d(x+c) \;\le\;
(4/3) \cdot 2^d \cdot {\rm e}^{2\sqrt{3}\,d} \;\le\;
(4/3) \cdot {\rm e}^{5d},$$
since~$c \le 1$.
Further, if~$C \le \size{x} \le 1$, we have~$(1-x^2)^{d_1} \le
{\rm e}^{-x^2 d_1} \le {\rm e}^{-6d}$.
Combining these two inequalities, we may bound~$r$ as follows: 
$$ \size{r(x)}  \eq  \size{\qq(x+c)} (1-x^2)^{d_1}
\;\le\; (4/3) \cdot {\rm e}^{5d} \cdot {\rm e}^{-6d} \;\le\; 1$$
for~$x$ in the region~$[-1,-C]\union[C,1]$.
\end{proof}

We now turn to the remaining case.

{\bf Case~(b).} $c > 1-C$. Without loss of generality, we assume
that~$\Delta_\ell \le \ell',\ell \le n - \Delta_\ell$ (otherwise,
the bound we seek follows from Lemma~\ref{thm-first} above, 
since~$\sqrt{m(n-m)}/\Delta_\ell \le \sqrt{n/\Delta_\ell}\,$).
This implies, in particular, that~$c < 1$. Let~$\alpha_c = \cos^{-1} c$.
Since~$0.99 < 1-C < c < 1$, we have~$0 < \alpha_c < 1/4$.

We prove a degree lower bound for a {\em trigonometric
polynomial\/}~$s$ derived from~$\qq$. The polynomial~$s$ is
defined as:
$$ s(\theta) \eq \qq(\cos \theta) [\cos(d_1(\theta-\alpha_c))]^{d_2}, $$
where~$d_1 = \floor{1/(2\alpha_c)}$ and~$d_2 = c_1 \ceil{d/d_1}$, for
some integer constant~$c_1 \ge 1$ to be specified later. Let~$D$ be
the degree of the polynomial~$s$.
\begin{claim}
$D = O(d)$.
\end{claim}
\begin{proof}
First, note that since~$\cos \theta \ge 1 - \theta^2/2$ for~$\theta \in
[0,\pi/2]$, we have
$$\alpha_c \;\ge\; 2\sqrt{1-\cos\alpha_c} \eq 2\sqrt{1-c}
\;\ge\; 2\sqrt{2\Delta_\ell/n}.$$
The last
inequality follows from the assumption that~$\ell \le n-\Delta_\ell$. So~$d_1
\le 1/(2\alpha_c) = O(\sqrt{n/\Delta_\ell}\,)$ which is~$O(d)$, by
Lemma~\ref{thm-first}. We may now bound~$D$ as follows:
$$D \;\le\; d + d_2 d_1 \eq d + c_1 \ceil{d/d_1} d_1
\;\le\; d+c_1(d+d_1).$$
So~$D = O(d)$.
\end{proof}

Thus, it suffices to prove a lower bound of~$\Omega(\sqrt{m(n-m)}/
\Delta_\ell)$ for~$D$, which we do next.

Let~$\alpha_i = \cos^{-1} a_i$, for~$i = 0,\ldots,n$.

Again, if~$\norm{s}$ is bounded by~$2$, we get the lower bound easily:
if~$c = b$, then~$\size{s(\alpha_c)} \ge 2$, and there is a point~$a_i$ at a
distance at most~$2/n$ to the left of~$c$ such that~$\size{s(\alpha_i)}
\le 4/3$. We therefore have, for some~$\alpha \in [\alpha_c,\alpha_i]$,
that~$\size{s'(\alpha)} \ge
(2/3)/(\alpha_i-\alpha_c)$. Moreover, by the Mean Value Theorem, we
have~$\alpha_i-\alpha_c = \size{\cos \alpha_i - \cos\alpha_c}/\sin
\hat{\alpha}$ for some~$\hat{\alpha} \in [\alpha_c,\alpha_i]$. Note
that 
$$\sin\hat{\alpha} \;\ge\; \sin\alpha_c \;\ge\; \sin\alpha_\ell \;\ge\;
\sin\alpha_m \eq \sqrt{1-a_m^2}.$$
Thus,~$\size{s'(\alpha)} \ge (2/3)\sqrt{1-a_m^2}/(2/n)$, which gives
us~$D = \Omega(\sqrt{m(n-m)}\,) = \Omega(\sqrt{m(n-m)}/\Delta_\ell)$, when
combined with Fact~\ref{thm-bern}, the Bernstein Inequality for
trigonometric polynomials.
If~$c = a_\ell$, we can similarly argue that~$D =
\Omega(\sqrt{m(n-m)}/\Delta_\ell)$.

We now examine the case when~$\norm{s} > 2$. Claim~\ref{thm-bd2} below
shows that~$\size{s(\theta)}$ is bounded by~$1$ when~$\theta \in
[-\pi,-\pi+\alpha_c/2] \union [-\alpha_c/2,\alpha_c/2] \union
[\pi-\alpha_c/2,\pi]$. We may assume that the
point where the norm (which is greater than~$2$) 
is attained is in~$[0,\pi]$; the proof proceeds in an analogous manner
in the other case. This point is then close to
some point~$\alpha_i \in [\alpha_c/2,\pi-\alpha_c/2]$
where~$\size{s(\alpha_i)} \le 4/3$. Arguing as before, we get that, for
some points~$\alpha, \beta \in [\alpha_c/2,\pi-\alpha_c/2]$,~$
\size{s'(\alpha)} \ge \norm{s}(\sin \beta)/3(2/n)$. Further, 
$$\sin\beta \;\ge\; \sin {{\alpha_c}\over 2} \;\ge\;
{{\alpha_c}\over 4} \;\ge\;
{{\sin\alpha_c}\over 4} \;\ge\; {{\sin\alpha_m}\over 4}. $$
From Fact~\ref{thm-bern}, we
now get~$D = \Omega(\sqrt{m(n-m)}) = \Omega(\sqrt{m(n-m)}/\Delta_\ell)$.

We now prove that~$s$ is bounded in the region mentioned above.
\begin{claim}
\label{thm-bd2}
For all~$\theta \in [-\pi,-\pi+\alpha_c/2] \union
[-\alpha_c/2,\alpha_c/2] \union [\pi-\alpha_c/2,\pi]$, we
have~$\size{s(\theta)} \le 1$.
\end{claim}
\begin{proof}
We prove the claim for~$\theta \in [0,\alpha_c/2]$. The analysis
for~$\theta$ in the other intervals is similar (one exploits the fact
that~$\qq(\cos\theta)$ is an {\em even\/} function of~$\theta$, and that
the corollary to Fact~\ref{thm-chebbd} limits its behaviour
outside~$[\alpha_c,\pi-\alpha_c]$).

Let~$h(\theta) = [\cos(d_1(\theta-\alpha_c))]^{d_2}$. Then, for~$\theta
\in [0,\alpha_c]$,
$$ \size{h(\alpha_c-\theta)} \eq \size{\cos(d_1\theta)}^{d_2}
\;\le\; (1-(d_1\theta)^2/4)^{d_2} \;\le\; {\rm e}^{-d_2 (d_1\theta)^2/4}
\;\le\; {\rm e}^{-c_1 d \theta^2/(16\alpha_c)}.$$
The first inequality follows from the fact that~$\cos \phi \le
1-\phi^2/4$ for~$\phi \in [0,\pi/2]$ and that~$0\le d_1\alpha_c \le
1/2$. The second is a consequence of~$1+x \le {\rm e}^x$. The remaining
steps follow from the definitions of~$d_1, d_2$ and the fact
that~$\alpha_c \le 1/4$.

Further, Corollary~\ref{thm-cheb-cor} gives us the following bound on
the value of~$\qq$ outside the interval~$[-c,c]$:
$$ \size{\qq(c+x)} \;\le\; 2 \size{T_d(1+x/c)} \;\le\; 2\cdot {\rm
e}^{2d\sqrt{3x/c}} $$
for~$x \in [0,1-c]$. Since, for~$\theta \in [0,\alpha_c]$,
$$ \cos(\alpha_c-\theta) \eq \cos\alpha_c\,\cos\theta + \sin\alpha_c\,
\sin\theta \;\le\; \cos\alpha_c + \alpha_c\theta \eq c+\alpha_c\theta,$$
we have~$\size{\qq(\cos(\alpha_c-\theta))} \le 2\cdot {\rm e}^{2d\sqrt{ 3
\alpha_c\theta/c}} \le 2\cdot {\rm e}^{4d\sqrt{\alpha_c\theta}}$. So,
for~$\theta \in [0,\alpha_c/2]$,
$$ \size{s(\theta)} \eq \size{\qq(\cos(\alpha_c-(\alpha_c-\theta)))}
\size{h(\alpha_c-(\alpha_c-\theta))} \;\le\; 1, $$
provided~$c_1$ is chosen large enough (as may readily be verified, bearing
in mind that~$1/\alpha_c = O(d)$).
\end{proof}

This completes the derivation of the second lower bound on the degree~$d$ 
of the polynomial~$\qq$.
\end{proof}

Lemmas~\ref{thm-first} and~\ref{thm-second} together imply that~$d
= \Omega\left(\max\set{\sqrt{n/\Delta_\ell}, \sqrt{m(n-m)}/
\Delta_\ell}\right)$,
which is equivalent to the bound stated in Theorem~\ref{thm-main}.
\end{proofof}

\subsection{Applications to quantum black-box computation}
\label{sec-apps}

In this section, we use our degree lower bound in conjunction with
a result of Beals {\em et al.\/}~\cite{bbcmw} to derive lower
bounds for the quantum black-box complexity of approximating the
statistics defined in Section~\ref{sec-summ}.
The key lemma of~\cite{bbcmw} which we require is the following.
\begin{lemma}[Beals, Buhrman, Cleve, Mosca, de~Wolf]
\label{thm-bbcmw}
Let~$\cal A$ be a quantum algorithm that makes~$T$ calls to a boolean
oracle~$X$. Then, there is a real multilinear polynomial~$p(x_0,\ldots,
x_{n-1})$ of degree at most~$2T$ such that the acceptance probability
of~$\cal A$ on oracle input~$X = (x_0,\ldots,x_{n-1})$ is
exactly~$p(x_0,\ldots,x_{n-1})$.
\end{lemma}

We deduce Corollary~\ref{thm-q-main} from Theorem~\ref{thm-main} using
this lemma.

\begin{proofof}{Corollary~\ref{thm-q-main}}
Consider an oracle quantum algorithm~$\cal A$ that computes the
partial function~$f_{\ell,\ell'}$ with constant error
probability~$c < 1/2$ by making at most~$T$ oracle queries. From the
lemma above, we deduce that there is a multilinear
polynomial~$p(x_0,\ldots,x_{n-1})$ of degree at most~$2T$ that gives the
acceptance probability of~$\cal A$ with the oracle input~$X =
(x_0,\ldots,x_{n-1})$. Clearly,~$p$ approximates~$f_{\ell,\ell'}$ to
within~$c$:~$p(X) \ge 1-c$ when~$\size{X} = \ell$ and~$p(X) \leq c$
when~$\size{X} = \ell'$, and, moreover, the value of~$p(X)$ 
is restricted to the interval~$[0,1]$ for all~$X \in \set{0,1}^n$.
Theorem~\ref{thm-main} now immediately implies the result.
\end{proofof}
 
In the remainder of this section, we show how to reduce from partial
function computations of the type given in Corollary~\ref{thm-q-main}
to approximating the $k$th-smallest element and to approximate counting,
and show how bounds for approximating the median and the mean follow.
In this way, we are able to show new quantum query lower bounds for the
computation of these approximate statistics.

The following two lemmas specialize Corollary~\ref{thm-q-main} to cases
of interest to us. The first deals with functions~$f_{\ell,\ell'}$ such
that neither~$\ell'$ nor~$\ell$ is ``close'' to~$0$ or~$n$, and the
second covers the remaining case.

\begin{lemma}
\label{thm-delta1}
Let~$k,\Delta > 0$ be integers such that~$2\Delta < k < n-2\Delta$.
Then, the quantum query complexity of~$f_{k-\Delta,k+\Delta}$ is
$\Omega(\sqrt{n/\Delta} + \sqrt{k(n-k)}/\Delta)$.
\end{lemma}
\begin{proof}
We assume that~$k \leq n/2$; the other case is symmetric.
In applying Corollary~\ref{thm-q-main},~$\Delta_\ell = 2\Delta$.
Since~$k \leq n/2$,~$m=k-\Delta$. Moreover,~$(k-\Delta)(n-k+\Delta) >
(k/2)(n-k)$. Corollary~\ref{thm-q-main} now gives us the claimed bound.
\end{proof}

\begin{lemma}
\label{thm-delta2}
Let~$k,\Delta$ be integers such that~$0 < \Delta \le n/4$
and~$0\le k \le 2\Delta$.
Then, the quantum query complexity of~$f_{0,k+\Delta}$ is~$\Omega(
\sqrt{n/\Delta} + \sqrt{k(n-k)}/\Delta)$. The same bound holds
for~$f_{k-\Delta,n}$ if~$k \ge n-2\Delta$.
\end{lemma}
\begin{proof}
We prove the first part of the lemma; the other part follows by symmetry.
In applying Corollary~\ref{thm-q-main}, we have $\Delta_\ell = k+\Delta
\le 3\Delta$, and~$m=0$. Hence, we get a bound of~$\Omega(\sqrt{n/\Delta})$
for~$f_{0,k+\Delta}$.
For the lemma to hold, we need only show that the second term in the
claimed lower bound is of the order of the first
term:~$\sqrt{k(n-k)}/\Delta \leq \sqrt{(2\Delta)n}/\Delta =
O(\sqrt{n/\Delta})$.
\end{proof}

We now prove the rest of the lower bound theorems of
Section~\ref{sec-summ} by exhibiting reductions from suitable problems.
We first consider the problem of estimating the~$k$th-smallest element.

\begin{proofof}{Theorem~\ref{thm-kth-lb}}
We need only prove the bound when~$\Delta \le n/4$, since it holds
trivially otherwise. We assume that~$\Delta$ is integral. The same proof
works with~$\ceil{\Delta}$ substituted for~$\Delta$ for general~$\Delta$.

Note that the query complexity
of computing~$f_{\ell,\ell'}$ is the same as that of
computing~$f_{n-\ell,n-\ell'}$, since we may negate the oracle responses
in an algorithm for the former to get an algorithm for the latter, and
vice-versa. We now consider two cases:

{\bf Case~(a).} $2\Delta < k < n-2\Delta$. Any algorithm for computing
a~$\Delta$-approximate $k$th-smallest element clearly also
computes~$f_{n-k+\Delta,n-k-\Delta}$, and hence, by
Lemma~\ref{thm-delta1} and the observation above makes at
least~$\Omega(\sqrt{n/\Delta} + \sqrt{k(n-k)}/\Delta)$ queries.

{\bf Case~(b).} $k \leq 2\Delta$ or $k \geq n-2\Delta$.
If~$k \leq 2\Delta$, we reduce from the function $f_{n,n-k-\Delta}$ to
our problem.
Lemma~\ref{thm-delta2} along with the observation above now gives us the
required bound. Similarly, for~$k \geq n-2\Delta$, we reduce
from~$f_{n-k+\Delta,0}$ and get the bound.

This completes the proof of the theorem.
\end{proofof}

Since the problem of approximating the median is really a special case
of the more general problem of estimating the~$k$th-smallest element, we
get a lower bound for this problem as well.

\begin{proofof}{Corollary~\ref{thm-median-lb}}
For~$n$ odd, an~$\epsilon$-approximate median is a~$\Delta$-approximate~$k$th
smallest element for~$k = (n+1)/2$, and~$\Delta = \ceil{(\epsilon
n+1)/2}$. The lower bound of~$\Omega(1/\epsilon)$ now follows from
Theorem~\ref{thm-kth-lb}.
\end{proofof}

The lower bounds for estimating the median and the~$k$th-smallest
element continue to hold in the comparison tree model,
since any comparison between two input numbers (which is made by
querying a comparison oracle in this model) can be simulated by
making at most~$4$ queries to an oracle of the sort we consider above.

The proofs for the lower bounds for approximate counting is similar to
that of Theorem~\ref{thm-kth-lb} above; we only sketch them here.

\begin{proofof}{Theorem~\ref{thm-count-lb}}
We may assume that~$\Delta \le n/6$, since the lower bound is trivial
otherwise. Consider any algorithm that approximately counts to within an
additive error of~$\Delta$.
Fix any~$0 \le t \le n$. Suppose for any input~$X$
with~$\size{X} = t$, the algorithm outputs a~$\Delta$-approximate count
after~$T$ queries with probability at least~$2/3$. We then consider the
truncated version of the algorithm which stops after making~$T$ queries
and outputs~$1$ if the approximate count obtained (if any) lies in the
range~$(t-\Delta,t+\Delta)$ and~$0$ otherwise. Since the
original algorithm approximates to within~$\Delta$ for all inputs, the
truncated algorithm computes~$f_{t,t+\ceil{2\Delta}}$ and/or~$f_{t,t-
\ceil{2\Delta}}$
whenever these partial functions are well-defined (i.e., when~$t+2\Delta \le
n$ and/or~$t-2\Delta \ge 0$). Now, by considering the four cases~$t \le
4\Delta$,~$n-t\le 4\Delta$,~$4\Delta < t \le n/2$ and~$n/2 < t
< n-4\Delta$ separately, and reducing from a suitable partial function
(out of~$f_{t,t+\ceil{2\Delta}}$ and~$f_{t,t- \ceil{2\Delta}}$)
in each of these cases, we get the claimed lower bound.
\end{proofof}

Since the problem of approximate counting is a restriction of
the more general problem of estimating the mean of~$n$ numbers, the
lower bound for the latter problem
follows directly from Theorem~\ref{thm-count-lb}.

\begin{proofof}{Corollary~\ref{thm-mean-lb}}
If the input numbers are all~0/1, multiplying an~$\epsilon$-approximate
mean by~$n$ gives us an~$\epsilon n$-approximate count. From
Theorem~\ref{thm-count-lb}, we get that in the worst
case (i.e., when the number of ones in the input is~$\floor{n/2}$), the
number of queries required to solve the approximate mean
problem is~$\Omega(1/\epsilon)$.
\end{proofof}

Finally, we sketch the proof of the lower bound for approximate counting
to within some relative error.

\begin{proofof}{Theorem~\ref{thm-relcount-lb}}
To derive a lower bound for the number of queries~$T$ made to
approximate the number of ones for~$X$ such that~$t_X = t$, we consider
a truncated version of the algorithm obtained by running the algorithm
till it returns a value between~$(1-\epsilon)t$ and~$(1+\epsilon)t$ with
probability at least~$2/3$ for such inputs. Since the
algorithm correctly approximates the count to within a relative error
of~$\epsilon$ for {\em all\/} inputs, we can use it
to compute the functions~$f_{t,t+1}$, when~$\epsilon t \le 1/4$,
and~$f_{t',t}$, where~$t' =
\floor{(1-\epsilon)t/(1+\epsilon)}$, when~$1/4 < \epsilon t$.
Corollary~\ref{thm-q-main} now gives us the claimed bound.
\end{proofof}

\section{Some optimal or essentially optimal algorithms}
\label{sec-ub}

We now show that the quantum black-box bounds obtained in
the previous section are either tight or essentially tight by giving
algorithms for the problems for which no such (optimal or near optimal)
algorithm was known.

\subsection{An optimal distinguisher}
\label{sec-dist}

Recall the problem of computing the partial function~$f_{\ell,\ell'}$
defined in Section~\ref{sec-summ}. In this section, we show how 
this partial function may be computed optimally, i.e., within a
constant factor
of the lower bound of Corollary~\ref{thm-q-main}, thus proving
Theorem~\ref{thm-q-ub}. Along with
Lemma~\ref{thm-bbcmw}, this implies that the polynomial degree lower
bound we show in Theorem~\ref{thm-main} is within a constant factor of
the optimal, and hence that it is not possible to obtain better lower
bounds for the problems we consider using our technique.

Our algorithm actually computes the partial
function~$\hat{f}_{\ell,\ell'} : \set{0,1}^n \rightarrow \set{0,1}$,
where~$0\le \ell' < \ell \le n$, defined as:
$$
\hat{f}_{\ell,\ell'} \eq \left\{ \begin{array}{ll}
			      1 & \mbox{if }\size{X} \ge \ell \\
			      0 & \mbox{if }\size{X} \le \ell' \\
                                 \end{array}
                         \right.
$$
Clearly, any algorithm for this partial function also
computes~$f_{\ell,\ell'}$, and thus the lower bound for the latter also
holds this function. (To compute~$f_{\ell,\ell'}$ when~$\ell < \ell'$,
it suffices to compute~$f_{\ell',\ell}$ and negate the output.)

The algorithm~$D(X, \ell',\ell)$ for~$\hat{f}_{\ell,\ell'}$,
which we call a {\em distinguisher}, is, in fact, an immediate 
derivative of an approximate
counting algorithm of Brassard {\em et al.}~\cite{bht,mo,bhmt}, which
enables us to estimate the number of ones~$t_Y$ of a boolean function~$Y$
in a useful manner. 
\begin{theorem}[Brassard, H{\o}yer, Mosca, Tapp]
\label{thm-count}
There is a quantum black-box algorithm~$C(Y,P)$ that,
given oracle access to a
boolean function~$Y = (y_0,\ldots,y_{n-1})$,
and an explicit integer parameter~$P$, makes~$P$ calls to
the oracle~$Y$ and computes a number~$t\in [0,n]$ such that
$$ \size{t_Y - t} \;\le\; {\sqrt{t_Y(n-t_Y)}\over P} +
{\size{n-2t_Y}\over{4 P^2}} $$
with probability at least~$2/3$.
\end{theorem}

Let~$X$ be the input to the distinguisher~$D$, and
let~$m$ and~$\Delta_\ell$ be defined as in Section~\ref{sec-summ}.
Further, let~$P = \ceil{c(\sqrt{n/\Delta_\ell}  + \sqrt{m(n-m)}/\Delta_\ell)}$,
where~$c$ is a constant to be determined later, and let~$t =
C(X,P)$. The algorithm~$D(X,\ell',\ell)$
returns~$0$ if~$t < \ell'+\Delta_\ell/2$ and~$1$ otherwise. The
correctness of the algorithm follows from the claim below; its
optimality is clear from the choice of~$P$.
\begin{claim}
\label{thm-dist}
With probability at least~$2/3$,
if~$t_X \le \ell'$, then~$t < \ell' + \Delta_\ell/2$, and if~$t_X \ge
\ell$, then~$t> \ell'+\Delta_\ell/2$.
\end{claim}
We give the proof of this claim in Appendix~\ref{sec-proofs}.
We will see in the next section that this distinguishing capability
of~$D$ also
allows us to search for an element of a desired rank
nearly optimally.

\subsection{Approximating the~$k$th-smallest element}

Consider the problem of approximating the the~$k$th-smallest element in
the black-box model.
Recall that when provided
with a list~$X = (x_1,\ldots,x_{n-1})$ of numbers as an oracle,
and an explicit parameter~$\Delta > 1/2$, the task of is to
find an input number~$x_i$ (or the corresponding index~$i$)
such that~$x_i$ is
a~$j$th-smallest element for a~$j \in (k-\Delta,k+\Delta)$. Notice that
we may round~$\Delta$ to~$\ceil{\Delta}$ without changing the function
to be computed. We therefore assume that~$\Delta$ is an integer in the
sequel.

The description of the function to be computed in terms of {\em ranks\/}
of numbers in the input list needs to be
given carefully, since there may be repetition of numbers in the
list. To accommodate repetitions, we let $\rank(x_i)$ denote the {\em
set\/} of positions~$j \in \set{1,\ldots,n}$
at which~$x_i$ could occur when the list~$X$ is
arranged in non-decreasing order. A~$\Delta$-approximate~$k$th-smallest
element is thus a number~$x_i$ such that~$\rank(x_i) \intersect
(k-\Delta,k+\Delta)$ is non-empty.

In this section we give a near optimal 
quantum black-box algorithm for computing
a~$\Delta$-approximate~$k$th-smallest element.
No non-trivial algorithm was known for this problem for general~$k$.
Our algorithm is inspired by the minimum finding algorithm of D{\"u}rr and
H{\o}yer~\cite{dh}, and builds upon the general search algorithm
of Boyer {\it et al.}~\cite{bbht}
and the distinguisher of the last section obtained from the 
approximate counting algorithm of Brassard
{\it et al.}~\cite{bht,mo,bhmt}.
To compute an~$\epsilon$-approximate median within the bound stated in
Corollary~\ref{thm-median-ub}, one only has to run this algorithm with 
the parameters~$k$ and~$\Delta$ chosen appropriately.

\subsubsection*{An abstract algorithm}

We first present the skeleton of our algorithm using two hypothetical
procedures~$S(\cdot\, ,\cdot)$ and~$K(\cdot)$. For convenience, we
define~$x_{-1} = -\infty$, and~$x_n = \infty$.
The procedure~$S(i,j)$ returns an index chosen
uniformly at random from the set of indices~$l$ such that~$x_i < x_l
< x_j$, if such an index exists.
The procedure~$K(i)$ returns `yes' when~$x_i$
is a~$\Delta$-approximate $k$th-smallest element of~$X$, `$<$' if~$x$ has
rank at most~$k-\Delta$ (i.e.,~$\rank(x) \intersect (k-\Delta,n] \eq
\emptyset$) and `$>$' if~$x$ has rank at least~$k+\Delta$
(i.e.,~$\rank(x) \intersect [1,k+\Delta) \eq \emptyset$).
Our algorithm, which we refer to as~$\aaa(S,K)$, performs a binary
search on the list of input numbers with a random pivot 
using~$S$ and~$K$. It thus has the following form:
\begin{enumerate}
\item $i \leftarrow -1$,~$j \leftarrow n$.
\item $l \leftarrow S(i,j)$.
\item If~$K(l)$ returns `yes', output~$x_l$ (and/or~$l$) and stop.

      Else, if~$K(l)$ returns `$<$',~$i \leftarrow l$, go to step~2.

      Else, if~$K(l)$ returns `$>$',~$j \leftarrow l$, go to step~2.
\end{enumerate}
Call an execution of steps~2 and~3 a {\em stage}.
This algorithm always terminates and produces a correct solution
within~$n-2\Delta +2$ stages. However, the following lemma
tells us that the {\em expected\/} number of stages before termination
is small. Let~$N = \sqrt{n/\Delta}  + \sqrt{k(n-k)}/\Delta$.
\begin{lemma}
\label{thm-timebd}
The algorithm~$\aaa(S,K)$
terminates with success after an expected~$O(\log N)$
number of stages.
\end{lemma}
We defer the proof of this lemma to Appendix~\ref{sec-proofs}.
Note that the lemma guarantees that, with probability at least~$3/4$, 
the algorithm~$\aaa(S,K)$ terminates within~$O(\log N)$ stages.

We now consider the behaviour of the algorithm~$\aaa$ when the
(deterministic) procedure~$K(\cdot)$ is replaced by a randomized
subroutine~$K'(\cdot)$ with the following specification. On
input~$i$ (for some~$0 \le i < n$):
\begin{itemize}
\item if~$x_i$ is a~${\Delta\over 2}$-approximate $k$th-smallest
      element, output `yes';
\item else, if $\rank(x_i)$ is at most~$k-\Delta$, output~`$<$';
\item else, if $\rank(x_i)$ is at least~$k+\Delta$, output~`$>$';
\item else, if $\rank(x_i)$ is at least~$k-\Delta+1$ and at most~$k-\Delta/2$,
      probabilistically output either `yes' or~`$<$';
\item else, if $\rank(x_i)$ is at least~$k+\Delta/2$ and at
      most~$k+\Delta-1$, probabilistically output either `yes' or~`$>$'.
\end{itemize}

The algorithm~$\aaa(S,K')$ obtained by replacing the
subroutine~$K(\cdot)$ by~$K'(\cdot)$ clearly also always computes a
correct solution. Although it may require more iterations of steps~2
and~3 to arrive at a solution,
we show that the increase is by at most a constant factor.
\begin{lemma}
\label{thm-timebd2}
Let~$X$ be any input oracle. The expected number of
stages of the algorithm~$\aaa(S,K')$ with oracle~$X$
and parameter~$\Delta$ is at most the expected number of stages
of~$\aaa(S,K)$ on inputs~$X$ and~$\Delta/2$.
\end{lemma}
Appendix~\ref{sec-proofs} contains a proof of this lemma.
In light of Lemma~\ref{thm-timebd}, this implies that~$\aaa(S,K')$ 
also terminates after an expected~$O(\log N)$ number of stages.

Finally, we analyse the behaviour of the algorithm~$\aaa(S,K')$ when the
procedures~$S$ and~$K'$ are allowed to either report
failure or output an incorrect answer with some small probability. As
mentioned above, we may restrict the number of stages of the algorithm
to~$O(\log N)$ and yet achieve success with probability at least~$3/4$.
Now, if any of~$S$ or~$K'$ fails (or errs) with probability~$o(1/\log N)$,
the net probability of success will still be at least, say,~$2/3$.

\subsubsection*{A realization of the algorithm}

We are now ready to spell out the implementation of the two
procedures~$S$ and~$K'$ out of which the algorithm is built.

The subroutine~$S$ is derived from the generalized search algorithm of
Boyer {\em et al.}~\cite{bbht}, which enables us to sample uniformly
from the set of ones of a boolean function.
\begin{theorem}[Boyer, Brassard, H{\o}yer, Tapp]
\label{thm-sample}
There is a quantum black-box algorithm with access to 
a boolean oracle~$Y = (y_0,\ldots,y_{n-1})$
that makes~$O(\sqrt{n/t}\,)$ queries and returns an
index~$i$ chosen uniformly at random from the set~$\set{j\st y_j = 1}$
with probability at least~$2/3$ if~$\size{Y} \ge t$.
\end{theorem}
Note that the success probability of the procedure described
above may be amplified to~$1-2^{\Omega(T)}$ by repeating it at
most~$O(T)$ times, and returning a sample as soon as a `one' of~$Y$ is
obtained. It can easily be verified that a sample so generated  has
the uniform distribution over the ones of~$Y$. The procedure~$S(i,j)$ is
implemented by defining a boolean function~$Y = (y_0,\ldots,y_{n-1})$ 
by~$y_l = 1$ if and only if~$x_i < x_l < x_j$, and using the above sampling
procedure. Every time~$S$ is invoked in~$\aaa$, there are at
least~$\Omega(\Delta)$ ones in~$Y$, and hence this implementation meets
the targeted specification if the parameter~$t$ in
Theorem~\ref{thm-sample} is chosen to be~$\Omega(\Delta)$, and the
number of repetitions~$T$ of the sampler is chosen to
be~$\Theta(\log\log N)$.
Each ``query'' to the function~$Y$ requires two queries to the input
oracle~$X$. Our sampling procedure
thus makes~$O(\sqrt{n/\Delta}\,\log\log N)$ queries and succeeds with
probability~$1- o(1/\log N)$.

The subroutine~$K'(i)$ is implemented by using
the dintinguisher~$D$ of Section~\ref{sec-dist} to detect whether~$x_i$
has rank that is ``far'' from~$k$ or not, by looking at both, the number of
elements smaller, and the number of elements larger than it. 
The probability of correctness of~$D$ may be boosted
to~$1-2^{\Omega(T)}$ by repeating the algorithm~$O(T)$ times, and returning
the majority of the answers so obtained. We require that the 
probability of error of our implementation be~$o(1/\log N)$, so we take~$T$
to be~$\Theta(\log\log N)$.
The detailed description the implementation follows:
\begin{enumerate}

\item If~$k+\Delta-1> n$, go to step~2, otherwise continue. 
Let~$t_0 = \ceil{k+\Delta/2} -2$, and~$t_1 = k+\Delta-1$.
Note that~$0\le t_0 < t_1 \le n$, since~$k,\Delta \ge 1$.
Define a boolean function~$Y$ over a domain of size~$n$, with~$y_j
= 1$ if and only if~$x_j < x_i$.
If the distinguisher~$D(Y,t_0,t_1)$ returns `$0$', go to step~2.
Else, output `$>$'.

\item If~$k-\Delta < 0$, return `yes', otherwise continue.
Let~$t_0 = n-\floor{k-\Delta/2}-1$, and~$t_1 = n-k+\Delta$. Note
that we again have~$0 \le t_0 < t_1 \le n$.
Define a boolean function~$Y$ over a domain of size~$n$, with~$y_j=
1$ if and only if~$x_j > x_i$.
If the distinguisher~$D(Y,t_0,t_1)$ returns `$0$', output
`yes'. Else, output `$<$'.
\end{enumerate}
It is easy to verify that this meets the specification for~$K'$ with
probability~$1- o(1/\log N)$, and that it makes~$O(N\log\log N)$ queries
to the oracle~$X$.

By Lemma~\ref{thm-timebd2}, we conclude that the total number of queries
made to the oracle is~$O(N\log (N)\log\log N)$, as claimed in
Theorem~\ref{thm-kth-ub}. Observe that our implementation of~$S$
and~$K'$ uses only comparisons between the inputs numbers, and thus may
be adapted to work in the comparison tree model as well, with the same
bound on the number of oracle queries.

\subsection{Optimal approximate counting}

Recall from Section~\ref{sec-summ} 
that the problem of computing a~$\Delta$-approximate count
consists of computing a number in~$[0,n]$ which is within an additive
error of~$\Delta$ from the number of ones~$t_X$ of a given boolean oracle
input~$X = (x_0,\ldots,x_{n-1})$.

The algorithm we propose is entirely analogous to the {\em exact\/}
counting algorithm of Brassard {\it et al.\/}~\cite{bht,mo,bhmt}, and we
give only a sketch of it here. The algorithm consists of first invoking
the procedure~$C(X,P)$ of Theorem~\ref{thm-count} 
a few times (say, five times), with~$P =
\ceil{c\sqrt{n/\Delta}}$ (for some suitable constant~$c$), and getting
an estimate~$\tilde{t}$ by taking the median of the approximate counts
returned by~$C$. With high (constant) probability, this estimate is
within~$O(\min\set{t_X,n-t_X}+\Delta)$ of the actual count~$t_X$. The
algorithm then invokes~$C$ again, with~$P = \ceil{c_1 (\sqrt{n/\Delta} +
\sqrt{\tilde{t}(n-\tilde{t})}/\Delta )}$
(for a suitable constant~$c_1$) and outputs the value returned by~$C$.
It is easy to verify that with high (constant) probability, the approximate
count obtained is within the required range. An analysis similar to that
of the exact counting algorithm mentioned above yields the bound of
Theorem~\ref{thm-count-ub} on the expected number of queries made by the
algorithm.

\subsection*{Acknowledgements}

We would like to thank Lov Grover for stimulating discussions, Michele
Mosca for sending us a copy of~\cite{mo} and explaining the details of
the exact counting algorithm therein, and Umesh Vazirani 
for his guidance and useful suggestions.

\appendix

\section{Some properties of polynomials}
\label{sec-prop}

In this section, we present some properties of polynomials and define
some concepts that we will use for our results.

The {\em symmetrization\/}~$p^{\rm sym}$ of
a multivariate polynomial~$p(x_0, \ldots,
x_{n-1})$ is defined to be
$$ p^{\rm sym}(x_0,\ldots,x_{n-1}) \eq \frac{\sum_{\pi \in S_n}
     p(x_{\pi(0)},\ldots,x_{\pi(n-1)})}{n!}\quad ,  $$
where~$S_n$ is the set of permutations on~$n$ symbols.

If~$p$ is a multilinear polynomial of degree~$d$, then~$p^{\rm sym}$
is also a multilinear polynomial of degree~$d$. Clearly,~$p^{\rm
sym}$ is a {\em symmetric\/} function.
The following fact attributed to Minsky and
Papert~\cite{mp} says that there is a succint representation
for~$p^{\rm sym}$ as a {\em univariate\/} polynomial.

\begin{fact}
\label{thm-sym}
If~$p : R^n \rightarrow R$ is a multilinear polynomial of degree~$d$,
then there exists a polynomial~$q : R \rightarrow R$, of degree at
most~$d$, such that~$q(x_0+x_1+\cdots +x_{n-1}) = p^{\rm sym}
(x_0,\ldots, x_{n-1})$ for~$x_i \in \{0,1\}$.
\end{fact}

In the remainder of this section, we will deal only with univariate
polynomials over the reals.

The properties of polynomials that we use involve the concept
of the {\em uniform\/} or {\em
Chebyshev norm\/} of a polynomial (denoted by~$\norm{p}$, for a 
polynomial~$p$), which is defined as follows:
$ \norm{p} \; =  \; \max_{-1 \le x \le 1} \size{p(x)}$.
We will refer to the uniform norm of a polynomial as simply the
{\em norm\/} of the polynomial.

The first property we require is a bound on the value of a polynomial in
an interval, given a bound on its values at {\em integer\/} points in
the interval.
\begin{fact}
\label{thm-normbd}
Let~$p$ be a polynomial of degree~$d \le n$ such that~$\size{p(i)} \le
c$ for integers~$i = 0,\ldots,n$. Then~$\size{p(x)} \le 2^d \cdot c$
for all~$x $ in the interval~$[0,n]$.
\end{fact}
This fact follows easily from an examination of the {\em Lagrange
interpolation\/} for the polynomial~$p$; the details are omitted.

The next fact bounds the value of a polynomial {\em outside\/} the
interval~$[-1,1]$, in terms of its norm (i.e., its maximum value
{\em inside\/} the interval~$[-1,1]$).
Let~$T_d(x) = {1\over 2}[(x+\sqrt{x^2-1})^d + (x-\sqrt{x^2-1})^d]$. This
polynomial is known as the {\em Chebyshev polynomial\/} of degree~$d$.
Note that~$\size{T_d}$ is an {\em even\/} function of~$x$, and
that~$\size{T_d(1+x)} \le {\rm e}^{2\sqrt{2x+x^2}}$, for~$x \ge 0$.
\begin{fact}
\label{thm-chebbd}
Let~$p$ be a polynomial of degree at most~$d$. Then, for~$\size{x} > 1$,
$$\size{p(x)} \; \le \; \norm{p}\cdot \size{T_d(x)}. $$
\end{fact}
A proof of this fact may be found in Section~$2.7$ of~\cite{ri}.
We require an easy corollary of this fact.
\begin{corollary}
\label{thm-cheb-cor}
If~$p$ is a polynomial of degree at most~$d$ and~$\size{p(x)} \le
c $ for~$\size{x} \le a$, for some~$a > 0$, then
$$ \size{p(x)} \;\le \; c \size{T_d(x/a)} $$
for all~$x$ with~$\size{x} \ge a$.
\end{corollary}

At the heart of our lower bound proof is the following set of
inequalities, due to Bernstein and Markov, which relate the
size of the derivative~$p'$
of a polynomial~$p$ to the degree of~$p$. Proofs of these may be found 
in Section~3.4 of~\cite{pp} and Section~2.7 of~\cite{ri}.
\begin{fact}
\label{thm-bern-mark}
Let~$p$ be a polynomial of degree~$d$. Then, for~$x \in
[-1,1]$,
\begin{enumerate}
\item {\bf (Markov)}\ \ \ \ \
$ \size{p'(x)} \; \le \; d^2\norm{p}$;
\item {\bf (Bernstein)}\ \ \
$ \sqrt{1-x^2}\,\size{p'(x)} \; \le \; d\norm{p}$.
\end{enumerate}
\end{fact}

The next fact, which is a more general version of the Bernstein Inequality
for algebraic polynomials, deals
with {\em trigonometric polynomials}. A {\em trigonometric
polynomial\/}~$t(x)$ of degree~$d$ is a real linear combination of the
functions~$\cos ix$ and~$\sin ix$, where~$i$ is an integer in the
range~$[0,d]$. 
For a trigonometric polynomial~$t$, we define its norm to be~$\norm{t} = 
\max_{-\pi \le x \le \pi} \size{t(x)}$.
\begin{fact}
\label{thm-bern}
Let $t$ be a trigonometric polynomial of degree $d$. Then, for $x \in
[-\pi,\pi]$, $$ \size{t'(x)} \; \le \; d\norm{t}. $$
\end{fact}

\section{Proofs of some claims made in Section~\ref{sec-ub}}
\label{sec-proofs}

\begin{proofof}{Claim~\ref{thm-dist}}
Recall that~$m\in\set{\ell,\ell'}$ is such that~$\size{{n\over 2}-m}$ is
maximized, and that~$\ell' < \ell$. We prove the claim when~$m \le n/2$;
the analysis of the other case is symmetric and is omitted. If~$m\le n/2$,
then~$m = \ell'$. Theorem~\ref{thm-count} says that with probability at
least~$2/3$,
$$ \size{t_X - t} \;\le\; {\sqrt{t_X(n-t_X)}\over P} + {\size{n-2t_X}\over{4
P^2}}. $$
Then, if~$t_X \le \ell' = m \le n/2$, and if~$c$ is large enough,
\begin{eqnarray*}
\size{t-t_X} & < & \frac{\sqrt{\ell' n}}{c\sqrt{\ell'
                    n/2}/\Delta_\ell}  + \frac{n}{4(c^2
		    n/\Delta_\ell)} \\
	     & < & {{\Delta_\ell} \over 2}.
\end{eqnarray*}
So~$t \;<\; t_X + \Delta_\ell/2 \;\le\; \ell' + \Delta_\ell/2$.
At the same time, we also have~$t \ge g(t_X)$, where~$g(x)$ is the
function
$$ g(x) \eq x - {\sqrt{xn}\over P} - {{n}\over{4 P^2}}. $$
We show that~$g$ is an increasing function of~$x$ for~$x\ge \ell$ and
that~$g(\ell) > \ell-\Delta_\ell/2 = \ell'+\Delta_\ell/2$,
provided~$c$ is chosen large enough.

The derivative of~$g$,
$$ g'(x) \eq 1 - \frac{\sqrt{n}}{2P\sqrt{x}} $$
is an increasing function of~$x > 0$, and if~$c$ is large enough,
$$ g'(\ell) \;\ge\;
1 - \frac{\sqrt{n}}{2c\sqrt{n/\Delta_\ell}\sqrt{\ell}} \;>\; 0,
$$
since~$\ell \ge \Delta_\ell$. So~$g'(x) > 0$ for all~$x\ge \ell$,
and~$g$ is increasing for such~$x$. Moreover, if~$c$ is large enough, we have
\begin{enumerate}
\item ${n\over{4 P^2}} \;\le\; {n\over{4(c^2 n/\Delta_\ell)}} \;<\; 
{{\Delta_\ell}\over 4}$;

\item if~$\ell' > \Delta_\ell$, then~$\ell = \ell'+\Delta_\ell < 2\ell'$,
and~${\sqrt{\ell n}\over P} \;\le\; {\sqrt{2 \ell' n}\over {(c\sqrt{\ell'
n/2}/\Delta_\ell)}} \;<\; {{\Delta_\ell}\over 4}$; and

\item if~$\ell' \le \Delta_\ell$, then~$\ell \le 2\Delta_\ell$,
and~${\sqrt{\ell n}\over P} 
\;\le\; {\sqrt{2\Delta_\ell n}\over{(c\sqrt{n/\Delta_\ell})}} \;<\;
{{\Delta_\ell}\over 4}$.
\end{enumerate}
It follows from the observations made above, that
\begin{eqnarray*}
g(\ell) & = & \ell - {\sqrt{\ell n}\over P} 
                   - {{n}\over {4P^2}} \\
	& > & \ell - {{\Delta_\ell} \over 2},
\end{eqnarray*}
and~$t \ge g(t_X) > \ell - \Delta_\ell/2$ for
all~$X$ such that~$t_X \ge \ell$.

This completes the proof of the claim.
\end{proofof}

\begin{proofof}{Lemma~\ref{thm-timebd}}
We examine, for every number in the input list, the probability that it
is {\em ever\/} selected in step~2 of the algorithm. The expected number
of stages is the sum of these probabilities; we show
that this sum is~$O(\log N)$. We concentrate on the case when~$\Delta \le
k \le n-\Delta$. The analysis in the other cases is similar.

Consider any arrangement of the numbers in the input list in sorted
order. For~$-1 \le i_0 < k < j_0 \le n$, let~$p(l,i_0,j_0)$ denote the
probability that the index of the~$l$th number in the sorted list is {\em
ever\/} chosen in step~2 of the algorithm {\em after\/}~$i = i_0$ and~$j
= j_0$. We are interested in bounding~$p(l,-1,n)$ for each~$l$ in the
range~$[0,k-\Delta]\union[k+\Delta,n]$. (The sum of these probabilities
for~$l \in (k-\Delta,k+\Delta)$ is clearly~$1$.) Suppose~$l \le
k-\Delta$. We get the following recurrence by considering the result of
the first invocation of~$S$ after~$i = i_0, j=j_0$:
$$ p(l,i_0,j_0) \;\le\;
{1\over{j_0-i_0-1}}\left[1+\sum_{i_1=i_0+1}^{l-1} p(l,i_1,j_0) +
\sum_{j_1 = k+\Delta}^{j_0-1} p(l,i_0,j_1) \right].$$
(The inequality is due to the fact that there may be repetitions of
numbers in the input list.)
Furthermore,~$p(l,l-1,k+\Delta) \le
1/(k+\Delta-l)$. By induction, we now get
$$ p(l,i_0,j_0) \;\le\; {1\over{k+\Delta-l}} $$
for all~$-1\le i_0 < l \le k-\Delta$ and~$k+\Delta \le j_0 \le
n$. Similarly, when~$l \ge k+\Delta$, we get
$$ p(l,i_0,j_0) \;\le\; {1\over{l+\Delta-k}} $$
for all~$-1\le i_0 \le k-\Delta$ and~$k+\Delta \le l < j_0 \le n$. 
The expected number of stages is thus bounded by
$$ \sum_{l = 1}^{k-\Delta} {1\over{k+\Delta-l}} + 1 + \sum_{l =
k+\Delta}^{n} {1\over{l+\Delta-k}}, $$
which is at most
$$ \ln\frac{(k+\Delta-1)(n-k+\Delta)}{(2\Delta-1)^2} + 1 \;\le\;
\ln\frac{(2k)(2(n-k))}{\Delta^2} + 1 \eq O(\log N)$$
since~$\Delta \le k$ and~$\Delta \le n-k$, and~$\Delta \ge 1$.
This is the bound in the statement of the lemma.
\end{proofof}

\begin{proofof}{Lemma~\ref{thm-timebd2}}
Call a sequence of elements generated by some choice of random coin
tosses of the procedure~$S$ in an execution of the
algorithm~$\aaa(S,K)$ or~$\aaa(S,K')$ till termination, a {\em run}.
We compare runs of the
algorithm~$\aaa(S,K')$ with parameter~$\Delta$ with the runs of the
algorithm~$\aaa(S,K)$ with parameter~$\Delta/2$.
Observe that when we condition on a set of decisions~$D$ of~$K'$ for
every input index, each run of~$\aaa(S,K')$ is also a {\em prefix\/} of
runs of~$\aaa(S,K)$, that the sum of the probabilities of the
occurrence of the runs of~$\aaa(S,K)$
of which a particular run of~$\aaa(S,K')$ is a prefix, is equal to the
probability of the occurrence of that run
of~$\aaa(S,K')$, and, finally, that
exactly {\em one\/} prefix of any run of~$\aaa(S,K)$ is consistent with
the set of decisions~$D$ we condition on.
A straightforward calculation of the expected length of a
run of~$\aaa(S,K')$ now gives us the required bound.
\end{proofof}

\end{document}